\documentclass[showpacs,twocolumn,showkeys,preprintnumbers,amsmath,amssymb]{revtex4}

\usepackage{graphicx}
\usepackage{dcolumn}

\begin{document}
\title{A degenerate three-level laser with a parametric amplifier}
\author{Eyob Alebachew }
\email{yob_a@yahoo.com}
\author{K. Fesseha}
\affiliation{Department of Physics, Addis Ababa University, P. O.
Box 33085, Addis Ababa, Ethiopia}

\date{\today}

\begin{abstract}
The aim of this paper is to study the squeezing and statistical
properties of the light produced by a degenerate three-level laser
whose cavity contains a degenerate parametric amplifier. In this
quantum optical system the top and bottom levels of the
three-level atoms injected into the laser cavity are coupled by
the pump mode emerging from the parametric amplifier. For a linear
gain coefficient of 100 and for a cavity damping constant of 0.8,
the maximum intracavity squeezing is found at steady state and at
threshold to be $93\%$.
\end{abstract}



\maketitle
\section{Introduction}
There has been a considerable interest in the analysis of the
quantum properties of the squeezed light generated by various
quantum optical systems ~\cite{1,2,3,4,5,6,7,8,9}. In squeezed
light the fluctuations in one quadrature is below the vacuum level
at the expense of enhanced fluctuations in the other quadrature,
with the product of the uncertainties in the two quadratures
satisfying the uncertainty relation. In addition to exhibiting a
nonclassical feature, squeezed light has potential applications in
precision measurements and noiseless communications ~\cite{10,11}.

Some authors have studied the squeezing and statistical properties
of the light produced by three-level lasers when either the atoms
are initially prepared in a coherent superposition of the top and
bottom levels ~\cite{12,13,14} or when these levels are coupled by
a strong coherent light~\cite{13}. These studies show that a
three-level laser can under certain conditions generate squeezed
light. In such a laser, three-level atoms in a cascade
configuration are injected at a constant rate into the cavity
coupled to a vacuum reservoir via a single-port mirror. When a
three-level atom makes a transition from the top to bottom level
via the intermediate level, two photons are generated. The two
photons are highly correlated and this correlation is responsible
for the squeezing of the light produced by a three-level laser. On
the other hand, it is well known that a parametric oscillator is a
typical source of squeezed light ~\cite{2,3,4,5,6}, with a maximum
intracavity squeezing of $50\%$. Recently Fesseha ~\cite{12} has
studied a three-level laser with a parametric amplifier in which
three-level atoms, initially prepared in a coherent superposition
of the top and bottom levels, are injected into the cavity. He has
found that the effect of the parametric amplifier is to increase
the intracavity squeezing by a maximum of 50$\%$.

In this paper we consider a degenerate three-level laser whose
cavity contains a degenerate parametric amplifier (DPA) and
coupled to a vacuum reservoir. The top and bottom levels of the
three-level atoms injected into the cavity are coupled by the pump
mode emerging from the parametric amplifier. And the three-level
atoms are initially prepared in such a way that the probabilities
of finding the atoms at the top and bottom levels are equal. We
expect that a highly squeezed light can be generated by the
quantum optical system under consideration. Thus our interest is
to analyze the squeezing and statistical properties of the light
generated by this system.

We obtain, applying the master equation, stochastic differential
equations for the cavity mode variables associated with the normal
ordering. The solutions of the resulting equations are used to
determine the quadrature variance, the squeezing spectrum, and the
mean photon number.  Moreover, applying the same solutions, we
determine the antinormally ordered characteristic function with
the aid of which the $Q$ function is obtained. Then the $Q$
function is used to calculate the photon number distribution.

\section{stochastic differential equations}
Three-level atoms in a cascade configuration are injected into the
laser cavity at a constant rate $r_{a}$ and removed from the
cavity after a certain time $\tau$. We represent the top, middle,
and bottom levels of a three-level atom by $|a\rangle$,
$|b\rangle$ and $|c\rangle$, respectively. We assume the
transitions between levels $|a\rangle$ and $|b\rangle$ and between
levels $|b\rangle$ and $|c\rangle$ to be dipole allowed, with
direct transitions between levels $|a\rangle$ and $|c\rangle$ to
be dipole forbidden. We consider the case for which the cavity
mode is at resonance with the two transitions
$|a\rangle$$\rightarrow$ $|b\rangle$ and $|b\rangle$$\rightarrow$
$|c\rangle$ (see Fig. 1).

The Hamiltonian describing the coupling of levels $|a\rangle$ and
$|c\rangle$ by the pump mode emerging from the parametric
amplifier can be expressed as
\begin{equation}\label{1}
\hat H^{\prime}=i\frac{\Omega}{ 2}(|c\rangle\langle
a|-|a\rangle\langle c|),
\end{equation}
in which $\Omega=2g^{\prime}\mu$ with $g^{\prime}$ and $\mu$ being
respectively the coupling constant and the amplitude of the pump
mode. In addition, the interaction of a three-level atom with the
cavity mode can be described by the Hamiltonian
\begin{equation}\label{2}
\hat H^{\prime\prime}=ig[\hat a^{\dagger}(|b\rangle\langle
a|+|c\rangle \langle b|)-\hat a(|a\rangle\langle
b|+|b\rangle\langle c|)],
\end{equation}
 where $g$ is the coupling constant and
$\hat a$ is the annihilation operator for the cavity mode.
\begin{figure}
\includegraphics [height=4cm,angle=0]{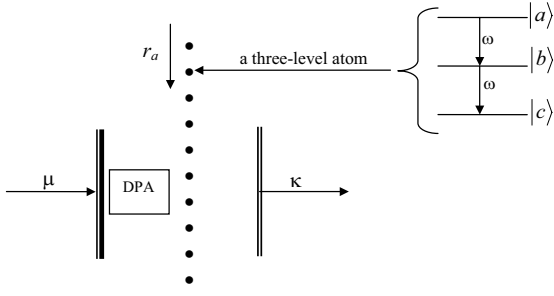}
\caption{A degenerate three-level laser with a degenerate
parametric amplifier.}
\end{figure}
Thus the Hamiltonian describing the interaction of a three-level
atom with the cavity mode and with the pump mode emerging from the
parametric amplifier has the form
\begin{align}\label{3}
\hat H &=ig[\hat a^{\dagger}(|b\rangle\langle a|+ |c\rangle
\langle
b|)-\hat a(|a\rangle\langle b|+|b\rangle\langle c|)]\notag\\
&+i\frac {\Omega} {2}(|c\rangle\langle a|-|a\rangle\langle c|).
\end{align}

We take the initial state of a single three-level atom to be
\begin{equation}\label{4}
|\psi_{A}(0)\rangle=\frac{1}{\sqrt{2}}|a\rangle+\frac{1}{\sqrt{2}}|c\rangle
\end{equation}
and hence the density operator for a single atom is
\begin{equation}\label{5}
\hat\rho_{A}(0)={1\over 2}|a\rangle\langle a|
+\frac{1}{2}|a\rangle\langle c|+{1\over 2}|c\rangle\langle
a|+{1\over 2}|c\rangle\langle c|.
\end{equation}
It can be readily established that the equation of evolution of
the density operator for the laser cavity mode, coupled to a
vacuum reservoir, has in the linear and adiabatic approximation
the form ~\cite{15}
\begin{align}\label{6}
\frac{d}{dt}\hat \rho&=R(2\hat a^{\dagger}\hat \rho\hat a-\hat
a\hat a^{\dagger}\hat \rho-\hat \rho\hat a\hat
a^{\dagger})\notag\\
&+S(2\hat a\hat \rho\hat a^{\dagger}-\hat
a^{\dagger}\hat a\hat \rho-\hat\rho\hat a^{\dagger}\hat a)\notag\\
&+U(\hat a^{\dagger}\hat \rho\hat a^{\dagger}+\hat a\hat\rho\hat
a-\hat\rho\hat a^{\dagger 2}-\hat a^2\hat \rho)\notag\\
&+V(\hat a^{\dagger}\hat \rho\hat a^{\dagger}+\hat a\hat\rho\hat
a-\rho\hat a^2-\hat a^{\dagger 2}\hat \rho),
\end{align} where
\begin{subequations}
\begin{equation}\label{7a}
R=\frac{A} {4B}\bigg[1-\frac{3\beta}{2}+\beta^2\bigg],
\end{equation}
\begin{equation}\label{7b}
S= \frac{A}{4B}\bigg[\frac{2\kappa
B}{A}+1+\frac{3\beta}{2}+\beta^2\bigg],
\end{equation}
\begin{equation}\label{7c}
U= \frac{A}{4B}\bigg[-1+{\beta\over 2}+{\beta^2\over
2}+{\beta^3\over 2} \bigg],
\end{equation}
\begin{equation}\label{7d}
V= {A\over 4B}\bigg[-1-{\beta\over 2}+{\beta^2\over
2}-{\beta^3\over 2} \bigg],
\end{equation}
\begin{equation}\label{7e}
B=(1+\beta^2)(1+{\beta^2\over 4}),
\end{equation}
\begin{equation}\label{7f}
\beta=\Omega/\gamma,
\end{equation}
\end{subequations}
\begin{equation}\label{8}
A={2g^2r_{a}\over \gamma^2}
\end{equation}
is the linear gain coefficient, $\kappa$ is the cavity damping
constant, and $\gamma$ is the atomic decay rate assumed to be the
same for all the three levels.

Moreover, a degenerate parametric amplifier with the pump mode
treated classically is describable in the interaction picture by
the Hamiltonian
\begin{equation}\label{9}
\hat H={i\varepsilon\over 2}(\hat a^{\dagger 2}-\hat a^2),
\end{equation}
in which $\varepsilon=\lambda\mu $ with $\lambda$ being the
coupling constant. The master equation associated with this
Hamiltonian has the form
\begin{equation}\label{10}
{d\over dt}\hat \rho={\varepsilon\over 2}(\hat \rho\hat a^2-\hat
a^2\hat\rho+\hat a^{\dagger 2}\hat\rho-\hat\rho\hat a^{\dagger
2}).\end{equation} Now on account of Eqs. \eqref{6} and
\eqref{10}, the master equation for the cavity mode of the quantum
optical system under consideration can be written as
\begin{align}\label{11}
\frac{d} {dt}\hat \rho &={\varepsilon\over 2}(\hat \rho\hat
a^2-\hat a^2\hat\rho+\hat a^{\dagger 2}\hat\rho-\hat\rho\hat
a^{\dagger 2})\notag\\
&+ R(2\hat a^{\dagger}\hat \rho\hat a-\hat a\hat a^{\dagger}\hat
\rho-\hat\rho\hat a\hat a^{\dagger})\notag\\
&+ S(2\hat a\hat \rho\hat a^{\dagger}-\hat a^{\dagger}\hat a\hat
\rho-\hat \rho\hat a^{\dagger}\hat a)\notag\\
&+U(\hat a^{\dagger}\hat \rho\hat a^{\dagger}+\hat a\hat\rho\hat
a-\hat\rho\hat a^{\dagger
2}-\hat a^2\hat \rho)\notag\\
&+ V(\hat a^{\dagger}\hat \rho\hat a^{\dagger}+\hat a\hat\rho\hat
a-\rho\hat a^2-\hat a^{\dagger 2}\hat \rho).
\end{align}

We next proceed to determine, applying this master equation, the
stochastic differential equations for the cavity mode variables.
To this end, applying \eqref{11} one readily finds
\begin{equation}\label{12}
{d\over dt}\langle \hat a\rangle=(R-S)\langle \hat a\rangle
+(U-V+\varepsilon)\langle \hat a^{\dagger}\rangle,
\end{equation}
\begin{equation}\label{13}
{d\over dt}\langle \hat a^2\rangle=2(R-S)\langle \hat a^2\rangle
+2(U-V+\varepsilon)\langle \hat a^{\dagger}\hat
a\rangle+\varepsilon-2V,
\end{equation}
\begin{equation}\label{14}
{d\over dt}\langle \hat a^{\dagger}\hat a\rangle=2(R-S)\langle
\hat a^{\dagger}\hat a\rangle +(U-V+\varepsilon)(\langle \hat
a^{\dagger 2}\rangle+\langle a^2\rangle)+2R.
\end{equation}
We note that these equations are in the normal order and the
corresponding c-number equations  are
\begin{equation}\label{15}
{d\over dt}\langle \alpha\rangle=-(S-R)\langle \alpha\rangle
+(U-V+\varepsilon)\langle \alpha^{*}\rangle,
\end{equation}
\begin{equation}\label{16}
{d\over dt}\langle \alpha^2\rangle=-2(S-R)\langle \alpha^2\rangle
+2(U-V+\varepsilon)\langle \alpha^{*}\alpha\rangle+\varepsilon-2V,
\end{equation}
\begin{align}\label{17}
{d\over dt}\langle \alpha^{*}\alpha\rangle &=-2(S-R)\langle
\alpha^{*}\alpha\rangle \notag\\
&+(U-V+\varepsilon)(\langle \alpha^{* 2}\rangle+\langle
\alpha^2\rangle)+2R.
\end{align}
On the basis of Eq. \eqref{15}, one can write the stochastic
differential equation
\begin{equation}\label{18}
{d\over dt}\alpha(t)=-(S-R)\alpha(t)
+(U-V+\varepsilon)\alpha^{*}(t)+f(t),
\end{equation}
where $f(t)$ is a noise force the properties of which remain to be
determined. We observe that Eq. \eqref{15} and the expectation
value of Eq. \eqref{18} will have the same form if
\begin{equation}\label{19}
\langle f(t)\rangle=0.
\end{equation}
Moreover, it can be readily verified using \eqref{18} that
\begin{align}\label{20}
{d\over dt}\langle \alpha^2(t)\rangle& =-2(S-R)\langle
\alpha^2(t)\rangle \notag\\
&+2(U-V+\varepsilon)\langle \alpha^{*}(t)\alpha(t)\rangle\notag\\
&+2\langle \alpha(t)f(t)\rangle,
\end{align}
and
\begin{align}\label{21}
{d\over dt}\langle \alpha^{*}(t)\alpha(t)\rangle & =-2(S-R)\langle
\alpha^{*}(t)\alpha(t)\rangle \notag\\
&+(U-V+\varepsilon)(\langle \alpha^{*
2}(t)\rangle+\langle \alpha^2(t)\rangle)\notag\\
&+ \langle \alpha(t)f^{*}(t)\rangle+\langle \alpha^{*}(t)
f(t)\rangle.
\end{align}
Comparison of  Eqs. \eqref{16} and \eqref{20} as well as Eqs.
\eqref{17} and \eqref{21} shows that
\begin{equation}\label{22}
\langle \alpha(t)f(t)\rangle={1\over 2}(\varepsilon -2V),
\end{equation}
\begin{equation}\label{23}\langle \alpha(t)f^{*}(t)\rangle+\langle
\alpha^{*}(t) f(t)\rangle=2R.
\end{equation}

Furthermore, a formal solution of Eq. \eqref{18} can be written as
\begin{align}\label{24}
\alpha(t)&=\alpha(0)e^{-(S-R)t}\notag\\
&+\int_{0}^t
e^{-(S-R)(t-t^{\prime})}[(U-V+\varepsilon)\alpha^{*}(t^{\prime})+f(t^{\prime})]dt^{\prime}.
\end{align}
Using Eq. \eqref{24} along with \eqref{22}, one easily finds
\begin{equation}\label{25}
\int_{0}^t e^{-(S-R)(t-t^{\prime})}\langle
f(t^{\prime})f(t)\rangle dt^{\prime}={1\over 2}(\varepsilon
-2V).\end{equation} Based on this result, one can write [12,15]
\begin{equation}\label{26}
\langle f(t^{\prime})f(t)\rangle=(\varepsilon
-2V)\delta(t-t^{\prime}).
\end{equation}
It can also be established in a similar manner that
\begin{equation}\label{27}
\langle f(t)f^{*}(t^{\prime})\rangle=2R\delta(t-t^{\prime}).
\end{equation}
We note that Eqs. \eqref{26} and \eqref{27} describe the
correlation properties of the noise force $f(t)$ associated with
the normal ordering.

Now introducing a new variable defined by
\begin{equation}\label{28}
\alpha_{\pm}(t)=\alpha^{*}(t)\pm\alpha(t),
\end{equation}
we easily get with the help of (18) that
\begin{equation}\label{29}
{d\over
dt}\alpha_{\pm}(t)=-\lambda_{\mp}\alpha_{\pm}(t)+f^{*}(t)\pm
f(t),\end{equation} where
\begin{equation}\label{30}
\lambda_{\mp}=(S-R)\mp(U-V+\varepsilon).
\end{equation}
The solution of Eq. \eqref{29} can be written as
\begin{equation}\label{31}
\alpha_{\pm}(t)=\alpha_{\pm}(0)e^{-\lambda_{\mp}t}+\int_{0}^{t}e^{-\lambda_{\mp}(t-t^{\prime})}(f^{*}(t^{\prime})\pm
f(t^{\prime}))dt^{\prime}.
\end{equation}
It then follows that
\begin{subequations}\label{32}
\begin{equation}\label{32a}
\alpha(t)=A(t)\alpha(0)+B(t)\alpha^{*}(0)+F(t),
\end{equation}
in which
\begin{equation}\label{32b}
A(t)={1\over 2}(e^{-\lambda_{-}t}+e^{-\lambda_{+}t}),
\end{equation}
\begin{equation}\label{32c}
B(t)={1\over 2}(e^{-\lambda_{-}t}-e^{-\lambda_{+}t}),
\end{equation}
\end{subequations}
and
\begin{subequations}\label{33}
\begin{equation}\label{33a}
F(t)=F_{+}(t)+F_{-}(t),
\end{equation}
with
\begin{equation}\label{33b}
F_{\pm}(t)={1\over
2}\int_{0}^{t}e^{-\lambda_{\mp}(t-t^{\prime})}(f^{*}(t^{\prime})\pm
f(t^{\prime}))dt^{\prime}.
\end{equation}
\end{subequations}

\section{Quadrature Fluctuations}

In this section we seek to calculate the quadrature variance and
squeezing spectrum for the cavity mode under consideration.

\subsection{Quadrature variance}

The variance of the quadrature operators
\begin{equation}\label{34}
\hat a_{+}=\hat a^{\dagger}+\hat a
\end{equation}
and
\begin{equation}\label{35}
\hat a_{-}=i(\hat a^{\dagger}-\hat a)
\end{equation}
is expressible in terms of c-number variables associated with the
normal ordering as
\begin{equation}\label{36}
\Delta
a_{\pm}^2=1\pm\langle\alpha_{\pm}(t),\alpha_{\pm}(t)\rangle,
\end{equation}
in which $\alpha_{\pm}(t)$ is given by Eq. \eqref{28}. Assuming
the cavity mode to be initially in a vacuum state and taking into
account \eqref{31} together with \eqref{19}, wee see that
\begin{equation}\label{37}
\langle\alpha_{\pm}(t)\rangle=0.
\end{equation}
In view of this result, Eq. \eqref{36} reduces to
\begin{equation}\label{38}
\Delta a_{\pm}^2=1\pm\langle\alpha_{\pm}^2(t)\rangle.
\end{equation}
Furthermore, employing Eq. \eqref{29}, one easily gets
\begin{align}\label{39}
{d\over dt}\langle\alpha_{\pm}^2(t)\rangle &=-2\lambda_{\mp}
\langle\alpha_{\pm}^2(t)\rangle+2\langle\alpha_{\pm}(t)f^{*}(t)\rangle\notag\\
&\pm 2\langle\alpha_{\pm}(t)f(t)\rangle.
\end{align}
With the aid of Eq. \eqref{31} along with \eqref{26} and
\eqref{27}, we readily obtain
\begin{equation}\label{40}
{d\over
dt}\langle\alpha_{\pm}^2(t)\rangle=-2\lambda_{\mp}\langle\alpha_{\pm}^2(t)\rangle+2(\varepsilon-2V\pm
2R).\end{equation} The steady-state solution of this equation
turns out to be
\begin{equation}\label{41}
\langle\alpha_{\pm}^2(t)\rangle={\varepsilon-2V\pm 2R\over
\lambda_{\mp}}.
\end{equation}
Now on account of \eqref{41} together with \eqref{30}, Eq.
\eqref{38} takes at steady state the form
\begin{figure}
\includegraphics [height=6.5cm,angle=0]{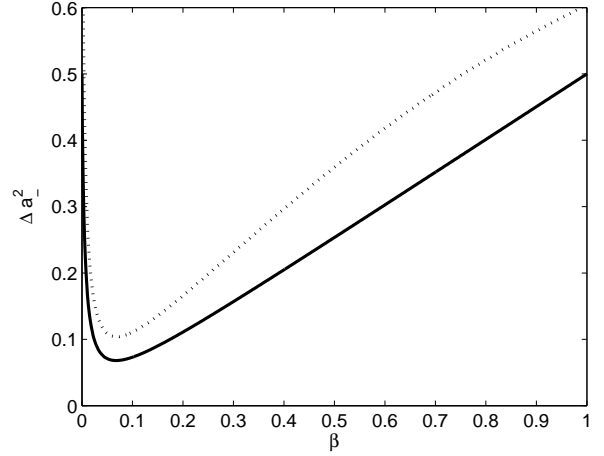}
\caption{Plots of the quadrature variance [Eq. \eqref{44b}] versus
$\beta$ for $\kappa=0.8$ and A=100 in the presence of the pump
mode and in the absence of the nonlinear crystal (dotted curve)
and [Eq. \eqref{46b}] for $\kappa=0.8$ and A=100 in the presence
of the parametric amplifier (solid curve).}
\end{figure}
\begin{figure}
\includegraphics [height=6.5cm,angle=0]{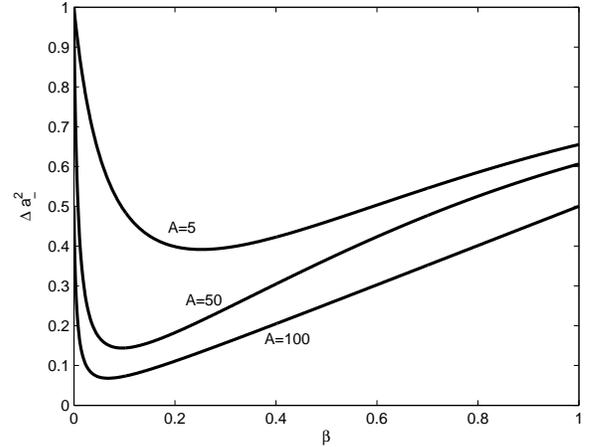}
\caption{Plots of the quadrature variance [Eq. \eqref{46b}] versus
$\beta$ for $\kappa=0.8$ and for different values of the linear
gain coefficient A in the presence of the parametric amplifier.}
\end{figure}
\begin{subequations}
\begin{equation}\label{42a}
\Delta a_{+}^2={ 2\kappa(1+\beta^2)(1+\beta^2/4)+A(4+\beta^2)
\over(2\kappa-4\varepsilon)(1+\beta^2)(1+\beta^2/4)+A(2\beta-\beta^3)}
\end{equation}
and
\begin{equation}\label{42b}
\Delta a_{-}^2={ 2\kappa(1+\beta^2)(1+\beta^2/4)+3A\beta^2 \over
(2\kappa+4\varepsilon)(1+\beta^2)(1+\beta^2/4)+A(4\beta+\beta^3)},
\end{equation}
\end{subequations}
where we have used Eqs. \eqref{7a}-\eqref{7e}.

In the absence of the parametric amplifier the quantum optical
system under consideration reduces to just a degenerate
three-level laser. The quadrature variance for this system has
upon setting $\beta=\varepsilon=0$ in \eqref{42a} and \eqref{42b}
the form
\begin{subequations}\label{43}
\begin{equation}
\Delta a_{+}^2={\kappa+2A\over\kappa}
\end{equation}
and
\begin{equation}
\Delta a_{-}^2=1.
\end{equation}
\end{subequations}
Since neither of the quadrature variance is less than one, the
light produced by the degenerate three-level laser is not in a
squeezed state. We therefore observe that the particular initial
preparation of the three-level atoms we have considered does not
lead to the generation of squeezed light. We next consider the
case in which three-level atoms are not injected into the cavity.
Thus setting $A=\beta=0$ (with $\mu\neq 0$) in Eqs. \eqref{42a}
and \eqref{42b}, we have
\begin{subequations}
\begin{equation}\label{43m1}
\Delta a_{+}^2={\kappa\over \kappa-2\varepsilon}
\end{equation}
and
\begin{equation}\label{43m2}
\Delta a_{-}^2={\kappa\over \kappa+2\varepsilon}.
\end{equation}
\end{subequations}
At threshold, $\varepsilon=\kappa/2$, the above equations reduce
to
\begin{subequations}
\begin{equation}\label{43m3}
\Delta a_{+}^2\rightarrow\infty
\end{equation}
and
\begin{equation}\label{43m4}
\Delta a_{-}^2=\frac{1}{2}.
\end{equation}
\end{subequations}
The squeezing in this case is exclusively due to the parametric
amplifier.

On the other hand, when the nonlinear crystal of the parametric
amplifier is removed from the cavity, the quantum optical system
under consideration reduces to a coherently driven three-level
laser. Hence upon setting $\varepsilon=\lambda\mu=0$ (with
$\mu\neq 0$) in Eqs. \eqref{42a} and \eqref{42b} the quadrature
variance for this system becomes
\begin{subequations}
\begin{equation}\label{44a}
\Delta a_{+}^2={ 2\kappa(1+\beta^2)(1+\beta^2/4)+A(4+\beta^2)
\over 2\kappa(1+\beta^2)(1+\beta^2/4)+A(2\beta-\beta^3)}
\end{equation}
and
\begin{equation}\label{44b}
\Delta a_{-}^2={ 2\kappa(1+\beta^2)(1+\beta^2/4)+3A\beta^2 \over
2\kappa(1+\beta^2)(1+\beta^2/4)+A(4\beta+\beta^3)}.
\end{equation}
\end{subequations}
It is interesting to consider the case for which the coherent
light driving the three-level laser is sufficiently strong. Thus
we note that for $\beta=\Omega/\gamma\gg 1$, Eq. \eqref{44a}
reduces to
\begin{equation}\label{44n1}
\Delta
a_{+}^2=\frac{\frac{\kappa}{2}+A/\beta^2}{\frac{\kappa}{2}-A/\beta}.
\end{equation}
Since $\beta^2$ is very large, we can drop the term $A/\beta^2$ in
Eq. \eqref{44n1}, so that
\begin{equation}\label{44n2}
\Delta a_{+}^2=\frac{\kappa}{\kappa-2A/\beta}.
\end{equation}
Following the same procedure, we also get
\begin{equation}\label{44n3}
\Delta a_{-}^2=\frac{\kappa}{\kappa+2A/\beta}.
\end{equation}
Now comparison of Eqs. \eqref{43m1} and \eqref{44n2} as well as
Eqs. \eqref{43m2} and \eqref{44n3} shows that a degenerate
three-level laser driven by a strong coherent light behaves in
exactly the same manner as a degenerate parametric oscillator,
which is in agreement with the assertion made in Ref.~\cite{16}.
However, as can be seen clearly from Fig. 2, a relatively high
degree of squeezing occurs for small values of $\beta$.

 Inspection of Eq. \eqref{30} shows that $\lambda_{+}$ is
nonnegative while $\lambda_{-}$ can be positive, negative or zero.
We thus note that Eq. \eqref{29} will not have a well-behaved
solution if $\lambda_{-}<0$.  Now setting $\lambda_{-}=0$ and
taking into account \eqref{30}, we get
\begin{equation}\label{45}
\varepsilon={\kappa\over 2}+{A(2\beta-\beta^3)\over
4(1+\beta^2)(1+\beta^2/4)}.
\end{equation}
We can then interpret this as the threshold condition for the
system under consideration. Therefore expressions \eqref{42a} and
\eqref{42b} take at threshold the form
\begin{subequations}\label{46}
\begin{equation}\label{46a}\Delta a_{+}^2\rightarrow \infty
\end{equation}
and
\begin{equation}\label{46b}
\Delta a_{-}^2={ 2\kappa(1+\beta^2)(1+\beta^2/4)+3A\beta^2 \over
4\kappa(1+\beta^2)(1+\beta^2/4)+6A\beta}.
\end{equation}
\end{subequations}
The dotted curve in Fig. 2 represents the quadrature variance for
the coherently driven three-level laser and the solid curve in the
same figure represents the quadrature variance for the degenerate
three-level laser with the parametric amplifier at threshold. It
is easy to see from Fig. 2 that the presence of the parametric
amplifier leads to better squeezing. Moreover, the minimum value
of the quadrature variance, described by Eq. \eqref{46b}, for
A=100 and $\kappa=0.8$ is found to be $\Delta a_{-}^{2}=0.068$ and
occurs at $\beta=0.067$. This result implies that the maximum
intracavity squeezing for the above values of the linear gain
coefficient and cavity damping constant is 93$\%$ below the vacuum
level. In addition Fig. 3 shows that the degree of squeezing
increases with the linear gain coefficient.

\subsection{Squeezing spectrum}

The squeezing spectrum of a single-mode light is expressible in
terms of c-number variables associated with the normal ordering as
\begin{subequations}\label{47}
\begin{equation}\label{47a}
S_{\pm}^{out}(\omega)=1\pm
2Re\int_{0}^{\infty}\langle\alpha_{\pm}^{out}(t),\alpha_{\pm}^{out}(t+\tau)\rangle_{ss}e^{i\omega\tau}d\tau,
\end{equation} where
\begin{equation}\label{47b}
\alpha_{\pm}^{out}(t)=\alpha_{out}^{*}(t)\pm\alpha_{out}(t).
\end{equation}
\end{subequations}
For a cavity mode coupled to  a vacuum reservoir, the output and
intracavity variables are related by
\begin{equation}\label{48}
\alpha_{\pm}^{out}(t)=\sqrt{\kappa}\alpha_{\pm}(t).
\end{equation}
Now taking in to account Eqs. \eqref{37} and \eqref{48}, the
squeezing spectrum can be put in the form
\begin{equation}\label{49}
S_{\pm}^{out}(\omega)=1\pm 2\kappa
Re\int_{0}^{\infty}\langle\alpha_{\pm}(t)\alpha_{\pm}(t+\tau)\rangle_{ss}e^{i\omega\tau}d\tau.
\end{equation}
On the other hand, the solution of Eq. \eqref{29} can also be
written as
\begin{align}\label{50}
\alpha_{\pm}(t+\tau)&= \alpha_{\pm}(t)
e^{-\lambda_{\mp}\tau}+\int_{0}^{\tau}e^{-\lambda_{\mp}(\tau-\tau^{\prime})}(f^{*}(t+\tau^{\prime})\notag\\
&\pm f(t+\tau^{\prime}))d\tau^{\prime}.
\end{align}

\begin{figure}
\includegraphics [height=6.5cm,angle=0]{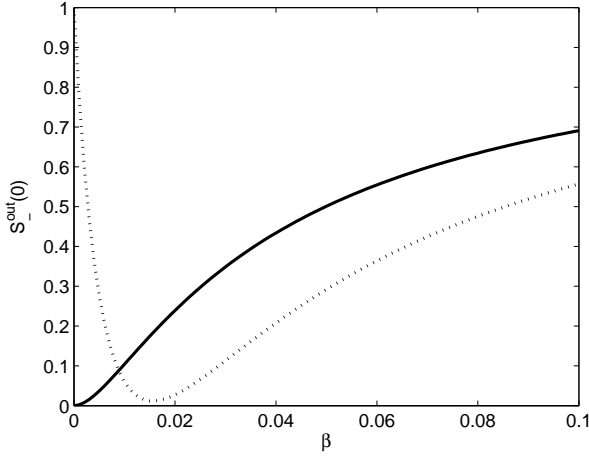}
\caption{Plots of the squeezing spectrum [Eq. \eqref{52b}] versus
$\beta$ for $\kappa=0.8$ and A=25 in the presence of the pump mode
and in the absence of the nonlinear crystal (dotted curve) and
[Eq. \eqref{53b}] for $\kappa=0.8$ and A=25 in the presence of the
parametric amplifier (solid curve).}
\end{figure}
\noindent Upon multiplying \eqref{50} by $\alpha_{\pm}(t)$ and
taking the expectation value of the resulting equation, one
readily obtains at steady state
\begin{equation}\label{51}
\langle\alpha_{\pm}(t)\alpha_{\pm}(t+\tau)\rangle_{ss}=
\langle\alpha_{\pm}^2(t)\rangle_{ss} e^{-\lambda_{\mp}\tau}.
\end{equation}
Now using Eqs. \eqref{41} and \eqref{51}, the squeezing spectrum
is found to be
\begin{subequations}\label{52}
\begin{equation}\label{52a}
S_{+}^{out}(\omega)=1+ {2\kappa(\varepsilon-2V+ 2R)\over
\lambda_{-}^2+\omega^2}
\end{equation}
and
\begin{equation}\label{52b}
S_{-}^{out}(\omega)=1- {2\kappa(\varepsilon-2V- 2R)\over
\lambda_{+}^2+\omega^2}.
\end{equation}
\end{subequations}
Applying Eqs. \eqref{45}, \eqref{7a}-\eqref{7e}, and \eqref{30},
the squeezing spectrum \eqref{52} can be put at threshold in the
form
\begin{subequations}\label{53}
\begin{equation}\label{53a}
S_{+}^{out}(\omega)=1+\frac{\kappa (\kappa
+\frac{A(4+\beta^2)}{(1+\beta^2)(2+\beta^2/2)})}{\omega^2}
\end{equation}
and
\begin{equation}\label{53b}
S_{-}^{out}(\omega)=1-\frac{\kappa(\kappa+\frac{A\beta(3-3\beta/2)}{(1+\beta^2)(1+\beta^2/4)})}
{(\kappa+\frac{3A\beta}{(1+\beta^2)(2+\beta^2/2)})^2+\omega^2}.
\end{equation}
\end{subequations}
Inspection of  this equation shows that there is perfect squeezing
for $\beta=\omega=0$ and for any values of A and $\kappa$. The
dotted curve in Fig. 4 shows that there is almost perfect
squeezing at $\beta=0.016$ and the solid curve indicates that the
presence of the parametric amplifier leads to perfect squeezing at
$\beta=0$.
\section{photon statistics}
We now proceed to calculate the mean photon number and the photon
number distribution for the cavity under consideration.
\subsection{The mean photon number}
Applying Eq.~\eqref{32a} and its complex conjugate, the mean
photon number of the cavity mode, assumed to be initially in a
vacuum state, can be written as
\begin{equation}\label{54}
\langle\alpha^{*}\alpha\rangle=\langle F^{*}(t)F(t)\rangle.
\end{equation}
On account of Eqs. \eqref{33a} and \eqref{33b} together with the
correlation properties of the noise force $f(t)$, we readily
obtain
\begin{align}\label{55}
\langle F^{*}(t)F(t)\rangle &={2R-2V+\varepsilon\over
4\lambda_{-}}(1-e^{-2\lambda_{-}t })\notag\\
&+{2R+2V-\varepsilon\over 4\lambda_{+}}(1-e^{-2\lambda_{+}t
}).\end{align} In view of this result, the mean photon number
takes the form
\begin{align}\label{56}
\langle\alpha^{*}\alpha\rangle &
={2\varepsilon(1+\beta^2)(1+\beta^2/4)+A(2-\beta+\beta^2/2+\beta^3/2)\over
4(1+\beta^2)(1+\beta^2/4)(\kappa-2\varepsilon)+2A(2\beta-\beta^3)}\notag\\
&\times(1-e^{-2\lambda_{-}t})\notag\\
-&
{2\varepsilon(1+\beta^2)(1+\beta^2/4)+A(2\beta-3\beta^2/2+\beta^3/2)\over
4(1+\beta^2)(1+\beta^2/4)(\kappa+2\varepsilon)+2A(4\beta+\beta^3)}\notag\\
&\times(1-e^{-2\lambda_{+}t }).
\end{align}
\begin{figure}
\includegraphics [height=6.5cm,angle=0]{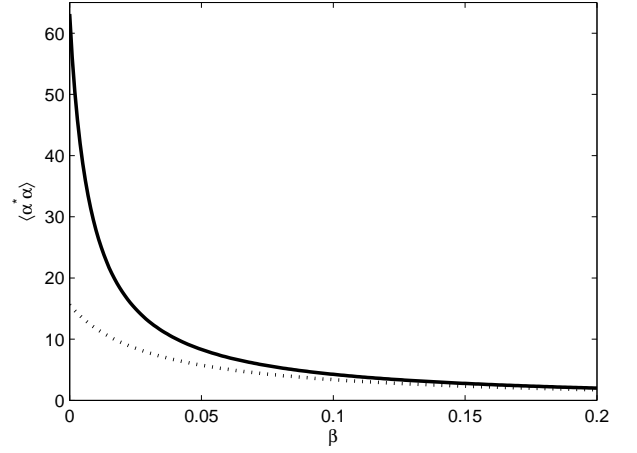}
\caption{Plots of the mean photon number [Eq. \eqref{56}] at
steady state versus $\beta$ for $\kappa=0.8$, and A=25 in the
presence of the pump mode and in the absence of the nonlinear
crystal (dotted curve) and for $\kappa=0.8$, and A=25
 in the presence of the parametric amplifier
with $\varepsilon =0.3$ (solid curve).}
\end{figure}

\noindent Fig. 5 clearly indicates that the parametric amplifier
contributes significantly to the mean photon number for relatively
small values of $\beta$.
\subsection{The Photon number distribution}
We finally seek to calculate, employing the $Q$ function, the
photon number distribution for the cavity mode. The photon number
distribution for a single-mode light is expressible in terms of
the $Q$ function as \cite{17}
\begin{equation}\label{57}
P(n,t)={\pi\over n!}{\partial^{2n}\over \partial \alpha^{*
n}\partial\alpha^n}
[Q(\alpha^{*},\alpha,t)e^{\alpha^{*}\alpha}]_{\alpha^{*}=\alpha=0}.
\end{equation}
Now using \eqref{57} and \eqref{A11}, the photon number
distribution for the cavity mode can be written in the form
\begin{align}\label{58}
P(n,t)&= {(c^2-d^2)^{1/2}\over n!}{\partial^{2n}\over \partial
\alpha^{* n}\partial\alpha^n}\notag \\
&\times exp[(1-c)\alpha^{*}\alpha-d(\alpha^{*
2}+\alpha^2)/2]_{\alpha^{*}=\alpha=0}.
\end{align}
Upon expanding the exponential functions in power series, we have
\begin{align}\label{59}
P(n,t)&= {(c^2-d^2)^{1/2}\over
n!}\sum_{klm}{(-1)^{l+m}(1-c)^kd^{l+m}\over 2^{l+m}k!l!m!}\notag\\
&\times{\partial^{2n}\over \partial \alpha^{*
n}\partial\alpha^n}[\alpha^{*
k+2l}\alpha^{k+2l}]_{\alpha^{*}=\alpha=0},
\end{align}
so that on carrying out the differentiations and applying the
condition ${\alpha^{*}=\alpha=0}$, there follows
\begin{align}\label{60}
P(n,t)&= {(c^2-d^2)^{1/2}\over
n!}\sum_{klm}{(-1)^{l+m}(1-c)^kd^{l+m}(k+2l)!\over
2^{l+m}k!l!m!(k+2l-n)!}\notag\\
&\times \frac{(k+2m)!}{(k+2m-n)!}\delta_{k+2l,n}\delta_{k+2m,n}.
\end{align}
Applying the properties of the Kronecker delta symbol and the fact
that a factorial is defined for nonnegative integers, we obtain
\begin{equation}\label{61}
P(n,t)= (c^2-d^2)^{1/2}\sum_{l=0}^{[n]}n!{(1-c)^{n-2l}d^{2l}\over
2^{2l}l!^2(n-2l)!},
\end{equation}
\begin{figure}[h]
\includegraphics[height=7.5cm,angle=0]{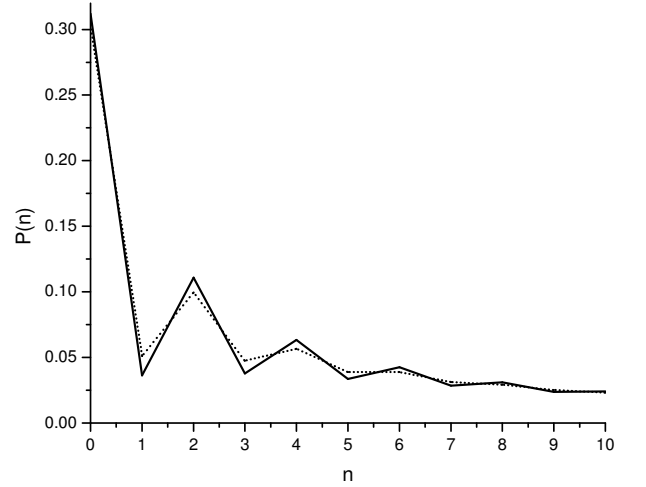}
\caption{Plots of the  photon number distribution
 [Eq.\eqref{61}] at steady state versus  photon number for A=100,
$\beta=0.067$, and $\kappa=0.8$ in the presence of the parametric
amplifier with $\varepsilon=0.3$ (solid curve), and
$\varepsilon=0$ in the presence of the pump mode and in the
absence of the nonlinear crystal (dotted curve).}
\end{figure}
where $[n]=n/2$ for even $n$ and $[n]=(n-1)/2$ for odd $n$. As can
be seen from Fig. 6, the steady-state photon number distribution
decreases with the photon number. Moreover, the probability of
finding even number of photons is in general greater than the
probability of finding odd number of photons. Although the photons
are generated in pairs in this quantum optical system, there is a
finite probability to find odd number of photons inside the
cavity. This is because some photons leave the cavity through the
port mirror. It also appears that the presence of the parametric
amplifier increases the probability of finding even number of
photons and decreases the probability of finding odd number of
photons.
\section{Conclusion}
In this paper we have considered a degenerate three-level laser
whose cavity contains a degenerate parametric amplifier, with the
top and bottom levels of the three-level atoms coupled by the pump
mode emerging from the parametric amplifier. We have obtained
using the master equation stochastic differential equations.
Applying the solutions of the resulting equations, we have
calculated the quadrature variance and squeezing spectrum.
Moreover, using the same solutions, we have determined the mean
photon number and the photon number distribution. We have found
that the light generated by the quantum optical system is in a
squeezed state, with the maximum intracavity squeezing attainable
being 93$\%$. The parametric amplifier increases the squeezing
significantly over and above the squeezing achievable due to the
coupling of the top and bottom levels by the pump mode. In
addition, we have seen that there is perfect squeezing of the
output light for $\beta=\omega=0$ and for any values of A and
$\kappa$. We have also found that the presence of the parametric
amplifier leads to a significant increase in the mean photon
number for small values of $\beta$. The plots of the photon number
distribution show that the probability of finding odd number of
photons is in general less than the probability of finding even
number of photons. Moreover, the same plots indicate that the
probability of finding even number of photons is greater in the
presence of the parametric amplifier than in the absence of the
nonlinear crystal. However, the opposite of this assertion holds
for the probability of finding odd number of photons.

\appendix
\section{The Q function}
We evaluate the $Q$ function applying the relation
\begin{equation}\label{A1}
Q(\alpha^{*},\alpha,t)={1\over \pi^2}\int
dz^2\phi(z^{*},z,t)exp(z^{*}\alpha-z\alpha^{*}),
\end{equation}
where
\begin{equation}\label{A2}
\phi(z^{*},z,t)=Tr(\hat\rho e^{-z^{*}\hat a(t)}e^{z\hat
a^{\dagger}(t)})
\end{equation}
is the antinormally ordered characteristic function defined in the
Hiesenberg picture. Using the identity
\begin{equation}\label{A3}
e^{\hat A}e^{\hat B}=e^{\hat B}e^{\hat A}e^{[\hat A,\hat B]},
\end{equation}
this function can be written in terms of c-number variables
associated with the normal ordering as
\begin{equation}\label{A4}
\phi(z^{*},z,t)=e^{-z^{*}z}\langle
exp(z\alpha^{*}(t)-z^{*}\alpha(t))\rangle.
\end{equation}
Since \eqref{15} is a linear differential equation, we see that
$\alpha(t)$ is a Gaussian variable~\cite{15}.  In addition, on
account of \eqref{32a}, \eqref{33a}, \eqref{33b}, and \eqref{19}
we easily find the mean value of $\alpha(t)$ to be zero.  Hence
$\alpha(t)$ is a Gaussian variable with vanishing mean.  One can
then express \eqref{A4} in the form ~\cite{18}
\begin{equation}\label{A5}
\phi(z^{*},z,t)= e^{-z^{*}z}exp ({1\over 2}\langle[z^2\alpha^{*
2}+z^{*2}\alpha^2-2zz^{*}\alpha\alpha^{*}]\rangle).
\end{equation}
Now employing Eqs. \eqref{32a}, \eqref{33a}, and \eqref{33b} along
with the correlation properties of the noise force $f(t)$, one can
establish that
\begin{align}\label{A6}
\langle \alpha^2\rangle &=\langle \alpha^{*
2}\rangle={2R-2V+\varepsilon\over
4\lambda_{-}}(1-e^{-2\lambda_{-}t })\notag\\
&-{2R+2V-\varepsilon\over 4\lambda_{+}}(1-e^{-2\lambda_{+}t
}),\end{align}
\begin{align}\label{A7}
\langle \alpha^{*}\alpha\rangle &={2R-2V+\varepsilon\over
4\lambda_{-}}(1-e^{-2\lambda_{-}t })\notag\\
&+{2R+2V-\varepsilon\over 4\lambda_{+}}(1-e^{-2\lambda_{+}t
}).\end{align} Therefore, with the aid of  Eqs. \eqref{A6} and
\eqref{A7}, Eq. \eqref{A5} can be put in the form
\begin{equation}\label{A8}
\phi(z^{*},z,t)=exp(-az^{*}z-b(z^2+z^{* 2})/2),
\end{equation}
in which
\begin{align}\label{A9}
a &=1+{2R-2V+\varepsilon\over 4\lambda_{-}}(1-e^{-2\lambda_{-}t
})\notag\\
&+{2R+2V-\varepsilon\over 4\lambda_{+}}(1-e^{-2\lambda_{+}t }),
\end{align}
\begin{align}\label{A10}
b &={2R-2V+\varepsilon\over 4\lambda_{-}}(1-e^{-2\lambda_{-}t
})\notag\\
&-{2R+2V-\varepsilon\over 4\lambda_{+}}(1-e^{-2\lambda_{+}t
}).\end{align} Upon substituting \eqref{A8} into \eqref{A1} and
performing the integration, the $Q$ function is found to be
\begin{equation}\label{A11}
Q(\alpha^{*},\alpha,t)={(c^2-d^2)^{1/2}\over
\pi}exp[-c\alpha^{*}\alpha-\frac{d}{2}(\alpha^{* 2}+\alpha^2)],
\end{equation}
where
\begin{equation}\label{A12}
c={a\over a^2-b^2},
\end{equation}
\begin{equation}\label{A13}
d={b\over a^2-b^2}.
\end{equation}

\end{document}